\documentclass[prd,onecolumn,nofootinbib,aps,superscriptaddress,tightenlines,preprintnumbers]{revtex4}

\usepackage{epsfig}
\usepackage{amsmath}

  \newcount\hour \newcount\minute
  \hour=\time \divide \hour by 60
  \minute=\time
  \newcommand{\mydate}{\ \today \ - \number\hour :\ifnum \minute<10 0\fi
\number\minute}

\def\nn{\nonumber \\ }

\newcommand{\pbra}[1]{\left(#1\right)}
\newcommand{\bbra}[1]{\left[#1\right]}

\newcommand{\abra}[1]{\left\langle#1\right\rangle}
\newcommand{\bfk}{{\bf k}}
\newcommand{\bfq}{{\bf q}}
\newcommand{\bfx}{{\bf x}}
\newcommand{\bfy}{{\bf y}}
\newcommand{\zstar}{{{\bf z}_*}}
\newcommand{\be}{\begin{equation}}
\newcommand{\ee}{\end{equation}}

\allowdisplaybreaks
\begin{document}

\preprint{\hbox{CALT-68-2705}  }

\title{Translational Invariance and the Anisotropy of the Cosmic Microwave Background}

\author{Sean M. Carroll, Chien-Yao Tseng and Mark B. Wise}

\affiliation{California Institute of Technology, Pasadena, CA 91125 }

\begin{abstract}
Primordial quantum fluctuations produced by inflation are conventionally assumed
to be statistically homogeneous, a consequence of translational invariance.
In this paper we quantify the potentially observable effects of a small violation of
translational invariance during inflation, as characterized by the
presence of a preferred point, line, or plane.  We explore the imprint such a violation
would leave on the cosmic microwave background anisotropy, and provide explicit
formulas for the expected amplitudes $\langle a_{lm}a_{l'm'}^*\rangle$ of the
spherical-harmonic coefficients.
\end{abstract}
\widetext
\date{\mydate}
\maketitle

\baselineskip=13pt

\section{Introduction}

Inflationary cosmology, originally proposed as a solution to the horizon, flatness, and monopole problems \cite{Guth:1980zm,{Linde:Albrecht}}, provides a very successful mechanism for generating primordial density perturbations.   During inflation, quantum vacuum fluctuations in a light scalar field are redshifted far outside the Hubble radius, imprinting an approximately scale-invariant spectrum of classical density perturbations \cite{quantumfluct,inflationreview}.  Models that realize this scenario have been widely discussed \cite{{Linde:1983gd},{Linde:1993cn},{Lyth:1998xn}}.  The resulting perturbations give rise to large-scale structure and temperature anisotropies in the cosmic microwave background, in excellent agreement with observation \cite{{COBE},{BOOMERANG},{ACBAR},{CBI},{VSA},{ARCHEOPS},{DASI},{MAXIMA},{WMAP3}}.

If density perturbations do arise from inflation, they provide a unique window on physics at otherwise inaccessible energy scales.  In a typical inflationary model (although certainly not in all of them), the amplitude of density fluctuations is of order $\delta\sim (E/M_{\rm P})^2$, where $E^4$ is the energy density during inflation and $M_{\rm P}$ is the (reduced) Planck mass.  Since we observe $\delta\sim 10^{-5}$, it is very plausible that inflation occurs near the scale of grand unification, and not too far from scales where quantum gravity is relevant.  Since direct experimental probes provide very few constraints on physics at such energies, it makes sense to be open-minded about what might happen during the inflationary era.

In a previous paper \cite{Ackerman}, henceforth ``ACW,'' the possibility that rotational invariance was violated by a small amount during the inflationary era was explored
(see also \cite{Gumrukcuoglu:2006xj,ArmendarizPicon:2007nr,Pereira:2007yy,Gumrukcuoglu:2007bx,Akofor:2007fv,Yokoyama:2008xw,Kanno:2008gn}). ACW suggested a simple, model-independent form for the power spectrum of fluctuations in the presence of a small violation of statistical isotropy, characterized by a preferred direction in space, and computed the imprint such a violation would leave on the anisotropy of the cosmic microwave background radiation. A toy model of a dynamical fixed-norm vector field  \cite{{Kostelecky:1989jw},{Jacobson:2000xp},{Jacobson:2004ts},{Carroll:2004ai},{Eling:2003rd},{Kostelecky:2003fs}} with a spacelike expectation value
was presented, which illustrated the validity of the model-independent arguments. The spacelike vector model is not fully realistic due to the presence of instabilities \cite{Himmetoglu}, and furthermore it does not provide a mechanism for turning off the violation of rotational invariance at the end of the inflationary era. Nevertheless, it still provides a useful check of the general argument that the terms which violate rotational invariance should be scale invariant. An inflationary era that violates rotational invariance results in a definite prediction, in terms of a few free parameters, for the deviation of the microwave background anisotropy that can be compared with the data \cite{Pullen:2007tu,Groeneboom:2008fz,ArmendarizPicon:2008yr}.

The results of ACW can be thought of as one step in a systematic exploration of
the ways in which inflationary perturbations could deviate by small amounts from the
standard picture, analogously to how the $STU$ parameters of particle physics \cite{PDG} parameterize
deviations from the Standard Model, or how the Parameterized Post-Newtonian (PPN)
formalism of gravity theory parameterizes deviations from general relativity \cite{Will}.
In cosmology, the fiducial model is characterized by primordial Gaussian perturbations
that are statistically homogeneous and isotropic,
with an approximately scale-free spectrum.  Even in the absence of an underlying
dynamical model, it is
useful to quantify how well existing and future experiments constrain departures from this
paradigm.  Deviations from a scale-free spectrum are quantified by the spectral index
$n_s$ and its derivatives; deviations from Gaussianity are quantified by the parameter
$f_{NL}$ of the three-point function (and its higher-order generalizations)
\cite{Linde:1996gt,Hu:2001fa,Acquaviva:2002ud,Maldacena:2002vr,Lyth:2005fi,Chen:2006nt}.
The remaining features of the fiducial model, statistical homogeneity and isotropy, are
derived from the spatial symmetries of the underlying dynamics.  

There is another important motivation for studying deviations from pure
statistical isotropy of cosmological perturbations:  a number of analyses have suggested
evidence that such deviations might exist in the real world \cite{Rakic:2007ve}.
These include the ``axis of evil'' alignment of low multipoles \cite{deOliveiraCosta:2003pu,
Schwarz:2004gk,Copi:2003kt,Land:2005ad,Gordon:2005ai,Land:2006bn,Copi:2006tu,Hajian:2007pi,Jain:1999},
the existence of an anomalous cold spot in the CMB \cite{Cruz:2004ce,Rudnick:2007kw,Smith:2008tc},
an anomalous dipole power asymmetry \cite{Eriksen:2003db,Hansen:2004vq,
Eriksen:2007pc,Gordon:2006ag,Erickcek:2008sm},
a claimed ``dark flow'' of galaxy clusters
measured by the Sunyaev-Zeldovich effect \cite{Kashlinsky:2008ut},
as well as a possible detection of a quadrupole power asymmetry of the type
predicted by ACW in the WMAP five-year data \cite{Groeneboom:2008fz}.
In none of these cases is it beyond a reasonable doubt that the effect is more than a
statistical fluctuation, or an unknown systematic effect; nevertheless, the combination of
all of them is suggestive \cite{Ralston:2004}.  It is possible that statistical isotropy/homogeneity is violated
at very high significance in some specific fashion that does not correspond precisely to any
of the particular observational effects that have been searched for, but that would stand
out dramatically in a better-targeted analysis.

The isometries of a flat Robertson-Walker cosmology are defined by  $E(3)$, the Euclidean group in three dimensions, which is generated by the three translations $R^3$ and the spatial rotations $O(3)$.  Our goal is to break as little of this symmetry as is possible in a consistent framework.  A preferred vector, considered by ACW \cite{Ackerman}, leaves all three translations unbroken, as well as an $O(2)$ representing rotations around the axis defined by the vector.  If we break some subgroup of the translations, there are three minimal possibilities, characterized by preferred Euclidean  submanifolds in space.  A preferred point breaks all of the translations, and preserves the entire rotational $O(3)$.  A preferred line leaves one translational generator unbroken, as well as one rotational generator around the axis defined by the line.  Finally, a preferred plane leaves the two translations within the plane unbroken, as well as a single rotation around an axis perpendicular to that plane.  We will consider each of these possibilities in this paper.

A random variable $\phi({\bf x})$ is statistically homogeneous (or translationally invariant) if all of its
correlation functions $\langle \phi({\bf x}_1)\phi({\bf x}_2) \cdots \rangle$ depend only
on the differences ${\bf x}_i - {\bf x}_j$, and is statistically isotropic (or rotationally
invariant) about some point ${\bf z}_*$ if the correlations depend only on
dot products of any of the vectors $({\bf x}_i - {\bf z}_*)$ and $({\bf x}_i - {\bf x}_j)$.  
The Fourier transform of the two-point function $\langle \phi({\bf x}_1)\phi({\bf x}_2) \rangle$
depends on two wavevectors ${\bf k}$ and ${\bf q}$, and will be translationally invariant if
it only has support when ${\bf k}= -{\bf q}$.  We will show how to perform a systematic expansion
in powers of ${\bf p} = {\bf k}+{\bf q}$.  ACW showed how
a small violation of rotational invariance during inflation would be manifested in
a violation of statistical isotropy of the CMB; here we perform a corresponding analysis
for a small violation of translational invariance.

At energies accessible to laboratory experiments, translational invariance plays a pivotal role, since it is responsible for the conservation of momentum. Here we are specifically concerned with the possibility that translational invariance may have been broken during inflation by an effect that disappeared after the inflationary era ended.  Such a phenomenon could conceivably arise from the presence of some sort of source that remained in our Hubble patch through inflation, although we do not consider any specific models along those lines.

\section{Setup For a Special Point}

In the standard inflationary cosmology the primordial density perturbations $ \delta({\bf x})$ have a Fourier transform ${\tilde \delta}({\bf k})$, defined by
\be
\delta ({\bf x})=\int d^3 k  e^{i {\bf k}\cdot {\bf x}} {\tilde  \delta} ({\bf k}),
\ee
and the power spectrum $P(k)$ is defined by
\begin{equation}
\label{powera}
\langle {\tilde {\delta}}({\bf k}){\tilde  {\delta}}({\bf q})\rangle=P(k) \delta^3({\bf k}+{\bf q}).
\end{equation}
so that
\be
\label{fourier}
\langle \delta ({\bf x}) \delta ({\bf y}) \rangle =\int d^3 k e^{i{\bf k}\cdot ({\bf x}-{\bf y)} }P(k).
\ee
The Dirac delta function in Eq.~(\ref{powera}) implies that modes with different wavenumbers are uncoupled.  This is a consequence of translational invariance during the inflationary era, while the fact that the power spectrum $P(k)$ only depends on the magnitude of the vector ${\bf k}$ is a consequence of rotational invariance.

Suppose that during the inflationary era translational invariance is broken by the presence of a special point with comoving coordinates ${\bf z_*}$.  This is reflected in the
statistical properties of the density perturbation $\delta({\bf x})$.  It is possible that the violation of translational invariance impacts the classical background for the inflation field during inflation and this induces a one point function,
\be
\langle \delta(\bfx)\rangle =G\left[|\bfx-\zstar |\right].
\ee
Throughout this paper we will assume that this classical piece is small (consistent with current
data) and concentrate on the
two-point function, which now takes the form
\be
\label{gen1}
  \langle \delta(\bfx)\delta(\bfy)\rangle = F\left[|\bfx-\bfy|, |\bfx-\zstar|, |\bfy-\zstar|,
  (\bfx-\zstar)\cdot(\bfy-\zstar)\right]\,,
\ee
where $F$ is symmetric under interchange of ${\bf x}$ and ${\bf y}$. This is the most general form 
of the two point correlation that is invariant under the transformations 
${\bf x} \rightarrow {\bf x}+{\bf a}$, ${\bf y} \rightarrow {\bf y}+{\bf a}$, 
${\bf \zstar} \rightarrow {\bf \zstar}+{\bf a}$, and rotational invariance about ${\bf z_*}$

It is convenient to  work with a form for  $\langle \delta(\bfx)\delta(\bfy)\rangle$   that is  analogous  to Eq.~(\ref{fourier}). We write,
\be
\label{gen2}
\langle \delta ({\bf x})  \delta({\bf y}) \rangle =\int d^3 k \int  d^3 q~ e^{i{\bf k}\cdot ({\bf x}-{\bf z_*})}e^{i{\bf q}\cdot ({\bf y}-{\bf z_*})}P_t(k, q, {\bf k} \cdot {\bf q}),
\ee
where $P_t$ is symmetric under interchange of $\bf k$ and $\bf q$. This is equivalent to Eq.~(\ref{gen1}) and is the most general form for the density perturbation's two-point correlation that breaks statistical translational invariance by the presence of a special point  ${\bf z}_*$, 
preserving rotational invariance about that point. In the limit where the violations of translational invariance are small and can be neglected,  the replacement $P_t(|{\bf k}|, |{\bf q}|, {\bf k} \cdot {\bf q}) \rightarrow P(k)\delta^3({\bf k}+{\bf q})$ is valid. 

We assume (as is consistent with the data)  that  violations of translational invariance are small and hence that $P_t$ is strongly peaked about ${\bf k}=-{\bf q}$. Hence we introduce the variables
${\bf p}={\bf k}+{\bf q}$, ${\bf l}=({\bf k}-{\bf q})/2$ and to expand in ${\bf p}$ using, for example,
\be
 k =| {\bf l} +{\bf p}/2 | = l +{{\bf p} \cdot {\bf l} \over 2l}-{({\bf p}\cdot {\bf l})^2 \over 8l^3}+{p^2 \over 8 l}+...
\ee
It is convenient to introduce $U_t={\rm ln}P_t$ and expand $U_t$ to quadratic order in ${\bf p}$,
neglecting the higher order terms since $P_t$ and hence $U_t$ is dominated by wavevectors 
${\bf p}$ near ${\bf p}=0$,
\be
\label{ptexpand}
P_t(|{\bf k}|, |{\bf q}|, {\bf k} \cdot {\bf q})=e^{U_t(l,l, -l^2)-A(l){p^2/2}-B(l)({\bf p} \cdot {\bf l})^2 /(2 l^2) +\ldots} \simeq P_t(l,l, -l^2)e^{-A(l){p^2}/2-B(l)({\bf p} \cdot {\bf l})^2 /(2 l^2)}.
\ee
Note that there are no terms linear in ${\bf p}$ because the symmetry under interchange of ${\bf k}$ and ${\bf q}$ implies symmetry under ${\bf l} \rightarrow  -{\bf l}$ and ${\bf p} \rightarrow  {\bf p}$.

Plugging the expansion of $P_t$ in Eq.~(\ref{ptexpand}) into Eq.~(\ref{gen2}) yields
\be
\langle \delta ({\bf x})  \delta({\bf y}) \rangle =\int d^3 l~  e^{i{\bf l}\cdot ({\bf x}-{\bf y})}P_t(l,l, -l^2)
\int d^3 p~  e^{-A(l){p^2}/2-B(l)({\bf p} \cdot {\bf l})^2 /(2l^2)} e^{i {\bf p}\cdot {\bf z}},
\ee
where ${\bf z}=({\bf x}+{\bf y}-2{\bf z_*})/2$. The integral over $ d^3 p$ can be performed by completing the square in the argument of the exponential. Introducing the $3\times 3$ matrix,
\be
C_{ij}=A(l)\delta_{ij}+B(l){ l_i l_j \over l^2}
\ee
we find that
\be
\int d^3 p~  e^{-A(l){p^2}/2-B(l)({\bf p} \cdot {\bf l})^2 /(2l^2)} e^{i {\bf p}\cdot {\bf z}}=\sqrt{{{(2\pi)^3} \over {\rm det }C}}e^{-z^TC^{-1}z/2}\simeq \sqrt{{{(2\pi)^3} \over {\rm det }C}}(1-z^TC^{-1}z/2).
\ee

Using this expression the two-point function can be written as
\be
\label{ptgen}
\langle \delta ({\bf x})  \delta({\bf y}) \rangle =\int d^3 l~  e^{i{\bf l}\cdot ({\bf x}-{\bf y})}P_t(l,l, -l^2)\sqrt{{{(2\pi)^3} \over {\rm det }C}}\left(1-{z^TC^{-1}z\over 2}+\ldots \right),
\ee
where the ellipses represent terms higher than quadratic order in the components of $\bf z$.  It is straightforward to solve for $C^{-1}$ and ${\rm det}C$ in terms of the functions $A$ and $B$ . We find that
${\rm det}C=A^3+A^2B$ and
\be
C^{-1}_{ij}={1 \over A}\delta_{ij}-{B \over A(A+B)}{l_il_j \over l^2}.
\ee
The part of the two point correlation that is rotationally invariant is the usual power spectrum $P(l)$, so
\be
\label{power0}
P(l)=\sqrt{{{(2\pi)^3} \over {\rm det }C}}P_t(l,l, -l^2).
\ee

Next we construct some mathematical examples that illustrate how the term proportional to $z^2$ is suppressed when $P_t$ is very strongly peaked at $p=0$. Without any  violation of translational invariance, $P_t(|{\bf k}|, |{\bf q}|, {\bf k} \cdot {\bf q}) = P(k)\delta^3({\bf k}+{\bf q})=c/k^3\delta^3({\bf k}+{\bf q})$ for a scale-invariant Harrison Zeldovich  power spectrum, where $c$ is some constant. We want to construct a form for $P_t$ that reduces to the standard Harrisson-Zeldovich spectrum with translational and rotational invariance as a parameter $d \rightarrow \infty$. The three dimensional delta function can be written as
\be
\delta^3({\bf k}+{\bf q})=\lim_{d\rightarrow\infty}\pbra{\frac{d}{\sqrt{\pi}}}^3e^{-d^2({\bf k}+{\bf q})^2}
\ee
Therefore, we might try writing $P_t$ as $c/k^3 \pbra{\frac{d}{\sqrt{\pi}}}^3e^{-d^2({\bf k}+{\bf q})^2}$ with $d$ a large number. However, this $P_t$ is not symmetric under the interchange of $\bf k$ and $\bf q$ because $k^3$ is not. 

There are many possible ways to resolve this problem. We might imagine replacing $k^3$ by 
$k^{3/2}q^{3/2}$,  $(k+q)^3/8$, $|{\bf k}-{\bf q}|^3/8$, $kq(k+q)/2$, $(k q)^{1/2}(k+q)^2/4$, $ ({\bf k}\cdot{\bf q})(k+q)/2$  $\cdots$,
or any linear combinations of these. With ${\bf p}={\bf k}+{\bf q}$, ${\bf l}=({\bf k}-{\bf q})/2$, we have
\be
k^{3/2}q^{3/2}=l^3\pbra{1-\frac{3({\bf p}\cdot {\bf l})^2}{4l^4}+\frac{3p^2}{8l^2}},
 \frac18(k+q)^3=l^3\pbra{1-\frac{3({\bf p}\cdot {\bf l})^2}{8l^4}+\frac{3p^2}{8l^2}}~~\cdots,
\ee
to second order in ${\bf p}$. Therefore, at quadratic order in ${\bf p}$,  the most general form of a function which is symmetric under the interchange of $\bf k$ and $\bf q$ and reduces to $k^3$ when $\bf k=-\bf q$ is
\be
l^3\pbra{1-a\frac{({\bf p}\cdot {\bf l})^2}{l^4}-b\frac{p^2}{l^2}},
\ee
with two parameters $a$ and $b$ that are independent of $l$. Hence we arrive at the following form for $P_t(|{\bf k}|, |{\bf q}|, {\bf k} \cdot {\bf q})$,
\be
P_t(|{\bf k}|, |{\bf q}|, {\bf k} \cdot {\bf q})=\frac{1}{l^3}c\pbra{1+a\frac{({\bf p}\cdot {\bf l})^2}{l^4}+b\frac{p^2}{l^2}}\pbra{\frac{d}{\sqrt{\pi}}}^3e^{-d^2p^2},
\ee
which gives the familiar translationally (and rotationally) invariant density perturbations with a Harrison-Zeldovich spectrum as $d \rightarrow \infty$.
Plugging into Eq.~\eqref{gen2}, the two-point function becomes
\be
\label{expan1}
\langle \delta ({\bf x})  \delta({\bf y}) \rangle =c(1-\frac{z^2}{4d^2})\int d^3l~e^{i{\bf l}\cdot ({\bf x}-{\bf y})}\frac{1}{l^3}\pbra{1+\frac{1}{2d^2}\frac{a+3b}{l^2}}.
\ee

We can construct another example which also gives dependence on $({\bf l}\cdot {\bf z})^2$.  First notice  that the three dimensional delta function can be written as another form,
\be
\delta^3({\bf p})=\lim_{d\rightarrow\infty}\pbra{\frac{d}{\sqrt{2\pi}}}^3\sqrt{{\rm det}U}e^{-\frac{d^2}{2}p^iU_{ij}p^j},
\ee
where $U_{ij}=2(\delta_{ij}+fl_il_j/l^2)$ and $f$ is an arbitrary parameter independent of $\bf l$. So another possible choice for $P_t$ that has the correct limiting behavior as $d \rightarrow \infty$ is
\be
P_t(|{\bf k}|, |{\bf q}|, {\bf k} \cdot {\bf q})=\frac{1}{l^3}c(1+a\frac{({\bf p}\cdot {\bf l})^2}{l^4}+b\frac{p^2}{l^2})\pbra{\frac{d}{\sqrt{2\pi}}}^3\sqrt{{\rm det}U}e^{-\frac{d^2}{2}p^iU_{ij}p^j}.
\ee
This gives,
\be
\label{expan2}
\langle \delta ({\bf x})  \delta({\bf y}) \rangle =\int d^3l~e^{i{\bf l}\cdot ({\bf x}-{\bf y})}\frac{1}{l^3}c\pbra{1+\frac{a+(3+2f)b}{2(1+f)d^2l^2}}\bbra{1-\frac{z^2}{4d^2}+\frac{f}{4(1+f)d^2}
\frac{({\bf l}\cdot {\bf z})^2}{l^2}}.
\ee

Since observable $|{\bf z}|$'s can be as large as our horizon, we need the parameter $d$ to be of that order (or larger) for the leading two terms of the expansion in $z$ to be a good approximation in Eq.~\eqref{expan1} and \eqref{expan2}.

The form we have derived in this section is plausible but is not the most general. For example, it could be that the Fourier transform of the two point function has the usual form plus a small piece that is proportional to a small parameter $\epsilon$. That is,
\be
\label{yikes}
P_t(|{\bf k}|, |{\bf q}|, {\bf k} \cdot {\bf q}) ={c \over k^3}\delta^3({\bf k}+{\bf q})+\epsilon P'_t(|{\bf k}|, |{\bf q}|, {\bf k} \cdot {\bf q})
\ee
If $\epsilon$ is small then the effects of the violation of translational invariance in Eq.~(\ref{yikes}) is small even when $P_t'$ is not strongly peaked about ${\bf k}=-{\bf q}$.

In the next section  we discuss how the violation of translational invariance during the inflationary era by the presence of a special point at fixed comoving coordinate impacts the anisotropy of the microwave background. Then in section IV we generalize the results of this section to the possibility that the violation of translation invariance during the inflationary era occurs because of a special line or plane during the inflationary era.
\section{Microwave Background Anisotropy with a Special Point}

We are interested in a quantitative understanding of how the second term in Eq.~\eqref{ptgen} changes the prediction for the microwave background asymmetry from the conventional translationally invariant one.  The multipole moments of the microwave background radiation are defined by
\begin{equation}
\label{def1}
a_{l m}=\int {\rm d} \Omega_{\bf e} Y_l^m({\bf e}){\Delta T \over T}({\bf e}).
\end{equation}
(Note that our definition differs from the conventional one\footnote{To shift our results  to what the the usual definition gives, $a_{l m} \rightarrow a_{lm}^*$. } in which the complex conjugate of $Y_l^m$ appears in the integral.)
Since the violation of translational invariance vanishes after the inflationary era ends, the anisotropy of the microwave background temperature $T$ along the direction of the unit vector $\bf e$ is related to the primordial fluctuations by
\begin{equation}
\label{def2}
{\Delta T \over T}({\bf e})=\int d^3k \sum_{l} \left({2l+1
\over 4 \pi}\right) (-i)^lP_l({\hat {\bf k}}\cdot {\bf e}){\tilde\delta} ({\bf k})\Theta_l(k),
\end{equation}
where $P_l$ is the Legendre polynomial of order $l$ and $\Theta_l(k)$ is a known real function of the magnitude of the wave vector ${\bf k}$ that includes, for example, the effects of the transfer function.

We are interested in computing $\langle a_{lm}a_{l'm'}^* \rangle$ to first order in the small correction that  violates translational invariance.  This is related to the two-point function in momentum space via
\be
\abra{a_{lm}a^*_{l'm'}}=(-i)^{l-l'}\int d^3k d^3q~Y_l^m(\hat \bfk){Y_{l'}^{m'}}^*(\hat \bfq)\Theta_l(k)\Theta_{l'}(q)\langle\tilde\delta(\bfk)\tilde\delta^*(\bfq)\rangle.
\ee
From Eq.~\eqref{ptgen} to Eq.\eqref{power0}, we have
\be
\label{ptgen1}
\langle \delta ({\bf x})  \delta({\bf y}) \rangle =\int d^3 l~ e^{i{\bf l}\cdot ({\bf x}-{\bf y})}P_0(l)+\frac{({\bf x}+{\bf y}-2{\bf z_*})^2}{4}\int d^3 l~ e^{i{\bf l}\cdot ({\bf x}-{\bf y})}P_1(l)+\int d^3 l~ e^{i{\bf l}\cdot ({\bf x}-{\bf y})}P_2(l)\frac{\bbra{\bf l\cdot({\bf x}+{\bf y}-2{\bf z_*})}^2}{4l^2}
\ee
where
\begin{eqnarray}
P_1(l)&=&-\frac{P_0(l)}{2A(l)}\\
P_2(l)&=&\frac{B(l)}{2A(l)\bbra{A(l)+B(l)}}P_0(l)
\end{eqnarray}
The models in Section II had $P_{1,2}(l)$ proportional to $P_0(l)$. The special point  ${\bf z_*}$ is characterized by three parameters; the magnitude of its distance from our location and two parameters for its direction (with respect to our location). Hence the corrections to the correlations  $\langle a_{lm}a_{l'm'}^* \rangle$ are characterized by just five parameters.
The Fourier transform of Eq.~\eqref{ptgen1} yields
\begin{eqnarray}
 \langle\tilde\delta(\bfk)\tilde\delta^*(\bfq)\rangle&=&\int {d^3x \over (2\pi)^3} \int {d^3y \over (2\pi)^3}~e^{-i\bfk\cdot\bfx}e^{i\bfq\cdot\bfy}
   \abra{\delta(\bfx)\delta(\bfy)}\nn
                            &=&P_{0}(k)\delta^3(\bfk-\bfq)+\frac{(i\nabla_\bfk-i\nabla_\bfq-2\zstar)^2}{4}P_{1}(k)\delta^3(\bfk-\bfq)\nn
                              &+&\sum_{i,j=1}^3 \frac14\pbra{i\frac{\partial}{\partial k_i}-i\frac{\partial}{\partial q_i}-2z_{*}^i}\pbra{i\frac{\partial}{\partial k_j}-i\frac{\partial}{\partial q_j}-2z_{*}^j}
                              \bbra{P_2(k)\frac{k_ik_j}{k^2}\delta^3(\bfk-\bfq)} .
\end{eqnarray}
We therefore define
\begin{equation}
  \langle a_{lm}a_{l'm'}^* \rangle=\langle a_{lm} a_{l'm'}^* \rangle_0+ (-i)^{l-l'}\Delta_1(l,m;l',m')+ (-i)^{l-l'}\Delta_2(l,m;l',m') ,
  \label{Delta}
\end{equation}
where the subscript $0$ denotes the usual translationally invariant piece,
\begin{equation}
\langle a_{lm} a_{l'm'}^* \rangle_0=\delta_{l,l'}\delta_{m, m'}
\int_0^{\infty} {\rm d}k k^2 P_{0}(k)\Theta_l(k)^2.
\end{equation}
and the correction coming from $P_1(k)$ is given by
\begin{eqnarray}
\Delta_1(l,m;l',m')&=&\frac14\int d^3kP_{1}(k)\left[-Y_l^m(\hat{\bf k})\Theta_l(k)\textrm{\boldmath $\nabla_k$}^2\pbra{Y_{l'}^{m'*}({\hat {\bf k}})\Theta_{l'}(k)}
-Y_{l'}^{m'*}(\hat{\bf k})\Theta_{l'}(k)\textrm{\boldmath $\nabla_k$}^2 \pbra{Y_{l}^{m}({\hat {\bf k}})\Theta_{l}(k)}\right.\nn
&&\left.+2\textrm{\boldmath $\nabla_k$} \pbra{Y_{l}^{m}({\hat {\bf k}})\Theta_{l}(k)}\cdot\textrm{\boldmath $\nabla_k$}\pbra{Y_{l'}^{m'*}({\hat {\bf k}})\Theta_{l'}(k)}
+4z_*^2~Y_l^m(\hat{\bf k})Y_{l'}^{m'*}(\hat{\bf k})\Theta_l(k)\Theta_{l'}(k)\right.\nn
&&\left.+4iY_{l'}^{m'*}({\hat {\bf k}})\Theta_{l'}(k)~\zstar\cdot\textrm{\boldmath $\nabla_k$} \pbra{Y_{l}^{m}({\hat {\bf k}})\Theta_{l}(k)}
-4iY_{l}^{m}({\hat {\bf k}})\Theta_{l}(k)~\zstar\cdot\textrm{\boldmath $\nabla_k$} \pbra{Y_{l'}^{m'*}({\hat {\bf k}})\Theta_{l'}(k)}\right].
\end{eqnarray}

It is convenient to break up $\Delta_1(l,m;l',m')$ into the parts quadratic in ${\bf z}_*$, linear in ${\bf z}_*$, and independent of ${\bf z}_*$, by writing
\begin{equation}
\Delta_1(l,m;l',m')=\Delta_1^{(2)}(l,m;l',m')+\Delta_1^{(1)}(l,m;l',m')+\Delta_1^{(0)}(l,m;l',m').
  \label{deltadecomp}
\end{equation}
The quadratic piece is relatively simple,
\begin{equation}
\Delta_1^{(2)}(l,m;l',m')=\delta_{l,l'}\delta_{m,m'}~z_*^2 \int_0^{\infty}{\rm d}k k^2 \Theta_l(k)^2 P_{1}(k).
  \label{delta2}
\end{equation}

The term linear in ${\bf z}_*$ is the most complicated. It can be evaluated using the identity
\begin{equation}
\label{tricky1}
i {\textrm{\boldmath $\nabla_k$}}(\Theta_l(k)Y_l^m({\hat{\bf k}}))=i\hat{\bf k}\left({\partial \Theta_l(k) \over \partial k} \right) Y_l^m({\hat{\bf k}})+{ 1 \over k} {\hat {\bf k}}{\bf \times}({\bf L}_{\bf k} Y_l^m({\hat{\bf k}})) \Theta_l(k),
\end{equation}
where ${ \bf L}_{\bf k}$ acts as the angular momentum operator in Fourier space,
\begin{equation}
\label{tricky}
{\bf L}_{\bf k} =-i {\bf k} {\textrm{\boldmath $\times$}} {\textrm{\boldmath $\nabla_k$}}.
\end{equation}
 It is convenient to divide $ \Delta_1^{(1)}(l,m;l',m')$ into a piece coming from the first term in Eq.~(\ref{tricky1}) and a term coming from the second term in Eq.~(\ref{tricky1}),
\begin{equation}
\label{linear}
\Delta_1^{(1)}(l,m;l',m')=\Delta_1^{(1)}(l,m;l',m')_a+\Delta_1^{(1)}(l,m;l',m')_b .
\end{equation}
To evaluate $\Delta_1^{(1)}(l,m;l',m')_{a,b}$, we express the components of ${\bf z}_*$ in terms of its ``spherical components,"
\begin{equation}
\label{sphericalzstar}
z_+=-{z_{*1}-i z_{*2}\over \sqrt{2}},~~~~z_-={z_{*1}+i z_{*2}\over \sqrt{2}},~~~~z_0=z_{*3} ,
\end{equation}
and express the components ${\bf \hat k}$ in terms of the spherical harmonics $Y_1^m({\bf{\hat k}})$. This gives
\begin{equation}
\Delta_1^{(1)}(l,m;l',m')_a=i\int_0^{\infty} {\rm d}k ~k^2 P_{1}(k)\left(\Theta_{l'}(k){\partial \Theta_l(k) \over \partial k}-\Theta_l(k) {\partial \Theta_{l'}(k) \over \partial k}\right)\left(z_+\chi^{(a)+}_{lm;l'm'} +z_-\chi^{(a)-}_{lm;l'm'} +z_0\chi^{(a)0}_{lm;l'm'} \right),
\label{termi}
\end{equation}
where
\begin{equation}
\label{chi}
\chi^{(a)0}_{l,m;l',m'}=\left[ (l-m+1)(l+m+1) \over (2l+1)(2l+3)\right]^{1/2}\delta_{l+1,l'}\delta_{m,m'}+\left[ (l-m)(l+m) \over (2l-1)(2l+1)\right]^{1/2}\delta_{l-1,l'}\delta_{m,m'},
\end{equation}
\begin{equation}
\chi^{(a)+}_{l,m;l',m'}={1 \over \sqrt 2}\left[ (l+m+1)(l+m+2) \over (2l+1)(2l+3)\right]^{1/2}\delta_{l+1,l'}\delta_{m+1,m'}-{1 \over \sqrt 2}\left[ (l-m)(l-m-1) \over (2l-1)(2l+1)\right]^{1/2}\delta_{l-1,l'}\delta_{m+1,m'},
\end{equation}
and
\begin{equation}
\label{chi-}
\chi^{(a)-}_{l,m;l',m'}=\chi^{(a)+}_{l,-m;l',-m'}.
\end{equation}
For $\Delta_1^{(1)}(l,m;l',m')_b$ we write
\begin{equation}
\Delta_1^{(1)}(l,m;l',m')_b= \Delta_1^{(1)'}(l,m;l',m')_b+{\Delta_1^{(1)'}(l',m';l,m)_b}^* ,
\end{equation}
and find that
\begin{equation}
\Delta_1^{(1)'}(l,m;l',m')_b=-i\int_0^{\infty}{\rm d }k~k P_{1}(k)\Theta_l(k)\Theta_{l'}(k)\left(z_+\chi^{(b)+}_{lm;l'm'} +z_-\chi^{(b)-}_{lm;l'm'} +z_0\chi^{(b)0}_{lm;l'm'} \right),
\end{equation}
where
\begin{equation}
\chi^{(b)0}_{l,m;l',m'}=l\left[ (l-m+1)(l+m+1)\over (2l+1)(2l+3)\right]^{1/2}\delta_{l+1,l'}\delta_{m,m'}-(l+1)\left[ (l-m)(l+m) \over (2l-1)(2l+1)\right]^{1/2}\delta_{l-1,l'}\delta_{m,m'},
\end{equation}
\begin{equation}
\chi^{(b)+}_{l,m;l',m'}={l \over \sqrt 2}\left[ (l+m+1)(l+m+2) \over (2l+1)(2l+3)\right]^{1/2}\delta_{l+1,l'}\delta_{m+1,m'}+{l+1 \over \sqrt 2}\left[ (l-m)(l-m-1) \over (2l-1)(2l+1)\right]^{1/2}\delta_{l-1,l'}\delta_{m+1,m'},
\end{equation}
and
\begin{equation}
\label{linear end}
\chi^{(b)-}_{l,m;l',m'}=\chi^{(b)+}_{l,-m;l',-m'}.
\end{equation}

Then we evaluate the term independent of ${\bf z}_*$ in $\Delta_1(l,m;l',m')$. Using integration by parts, we know
\begin{eqnarray}
\label{tricky2}
&&\int d^3kP_{1}(k)\textrm{\boldmath $\nabla_k$} \pbra{Y_{l}^{m}({\hat {\bf k}})\Theta_{l}(k)}\cdot\textrm{\boldmath $\nabla_k$}\pbra{Y_{l'}^{m'*}({\hat {\bf k}})\Theta_{l'}(k)}\nn
&&=\int d^3k\left[ -Y_{l'}^{m'*}({\hat {\bf k}})\Theta_{l'}(k)\frac{\partial P_1(k)}{\partial k}\hat{\bf k}\cdot\textrm{\boldmath $\nabla_k$} \pbra{Y_{l}^{m}({\hat {\bf k}})\Theta_{l}(k)}
-P_1(k)Y_{l'}^{m'*}({\hat {\bf k}})\Theta_{l'}(k)\textrm{\boldmath $\nabla_k$}^2 \pbra{Y_{l}^{m}({\hat {\bf k}})\Theta_{l}(k)}\right]
\end{eqnarray}
Another familiar result of spherical harmonics is
\begin{equation}
\label{tricky3}
-\nabla^2_{\bf k} Y_l^m({\hat{\bf k}}) \Theta_l(k)=\left[ -{ 1 \over k^2}{\partial \over \partial k }\left(k^2 {\partial \Theta_l(k)  \over \partial k} \right) +{l(l+1) \over k^2} \Theta_l(k) \right] Y_l^m({\hat{\bf k}}).
\end{equation}
Combining Eq.~\eqref{tricky1}, \eqref{tricky2}, and \eqref{tricky3} implies that,
\begin{equation}
\Delta_1^{(0)}(l,m;l',m')=\delta_{l,l'}\delta_{m,m'}\int_0^{\infty} {\rm d}k \left[- P_{1}(k)\Theta_l(k) { \partial \over \partial k} \left(k^2 {\partial \Theta_l(k) \over \partial k}\right) +l(l+1)P_{1}(k)\Theta_l(k)^2 -\frac12k^2\frac{\partial P_1(k)}{\partial k}\frac{\partial\Theta_l(k)}{\partial k}\Theta_l(k)\right]
  \label{delta0}
\end{equation}

The next step is to calculate the correction coming from $P_2(k)$.
\begin{eqnarray}
\Delta_2(l,m;l',m')&=&\frac14\int d^3kP_{2}(k)
\left[\vphantom{\sum_{i,j=1}^3\frac{k_ik_j}{k^2}}4(\hat{\bf k}\cdot\zstar)^2~Y_l^m(\hat{\bf k})Y_{l'}^{m'*}(\hat{\bf k})\Theta_l(k)\Theta_{l'}(k)\right.\nn
&&\left.+4i\pbra{\hat{\bf k}\cdot\zstar}\pbra{Y_{l'}^{m'*}({\hat {\bf k}})\Theta_{l'}(k)~\hat{\bf k}\cdot\textrm{\boldmath $\nabla_k$} \pbra{Y_{l}^{m}({\hat {\bf k}})\Theta_{l}(k)}
-Y_{l}^{m}({\hat {\bf k}})\Theta_{l}(k)~\hat{\bf k}\cdot\textrm{\boldmath $\nabla_k$} \pbra{Y_{l'}^{m'*}({\hat {\bf k}})\Theta_{l'}(k)}}\right.\nn
&&\left.-\sum_{i,j=1}^3\frac{k_ik_j}{k^2}\left(Y_l^m(\hat{\bf k})\Theta_l(k)\frac{\partial }{\partial k_i}\frac{\partial }{\partial k_j}\pbra{Y_{l'}^{m'*}({\hat {\bf k}})\Theta_{l'}(k)}
+Y_{l'}^{m'*}(\hat{\bf k})\Theta_{l'}(k)\frac{\partial }{\partial k_i}\frac{\partial }{\partial k_j} \pbra{Y_{l}^{m}({\hat {\bf k}})\Theta_{l}(k)}\right)\right.\nn
&&\left.\vphantom{\sum_{i,j=1}^3\frac{k_ik_j}{k^2}}+2~\pbra{\hat{\bf k}\cdot\textrm{\boldmath $\nabla_k$} \pbra{Y_{l}^{m}({\hat {\bf k}})\Theta_{l}(k)}}\pbra{\hat{\bf k}\cdot\textrm{\boldmath $\nabla_k$}\pbra{Y_{l'}^{m'*}({\hat {\bf k}})\Theta_{l'}(k)}}\right].
\end{eqnarray}

We also break ${\Delta_2}(l,m;l',m')$ into terms  quadratic in ${\bf z_*}$, linear and containing no factors of ${\bf z_*}$.
\begin{equation}
\label{deltadecomp2}
{\Delta_2}(l,m;l',m')={\Delta}_2^{(2)}(l,m;l',m')+{\Delta}_2^{(1)}(l,m;l',m')+{\Delta}_2^{(0)}(l,m;l',m')
\end{equation}
The term quadratic in ${\bf z_*}$ can be written as
\be
\label{delta22}
{\Delta}_2^{(2)}(l,m;l',m')=\xi_{lm;l'm'}\int_0^\infty{\rm d}kk^2P_{2}(k)\Theta_l(k)\Theta_{l'}(k)
\ee
where
\begin{equation}
\xi_{lm;l'm'}=\int{\rm d }\Omega_{\bf k} ({\hat {\bf k}}\cdot\zstar)^2Y_l^m(\hat{\bf k})Y_{l'}^{m'*}({\hat {\bf k}}),
\end{equation}
For the computation of $\xi_{l,m;l'm'}$, we use the ``spherical" components of $\zstar$ in Eq.~\eqref{sphericalzstar}.
$\xi_{lm;l'm'}$ was calculated in \cite{Ackerman} where violation of rotational invariance was considered. It is convenient to decompose
$\xi_{lm;l'm'}$ into coefficients of the quadratic quantities $z_iz_j$, via
\begin{equation}
\xi_{lm;l'm'}=z_+^2\xi_{lm;l'm'}^{++}+z_-^2\xi_{lm;l'm'}^{--}+2z_+z_-\xi_{lm;l'm'}^{+-}+2z_+z_0\xi_{lm;l'm'}^{+0}+2z_-z_0\xi_{lm;l'm'}^{-0}+z_0^2\xi_{lm;l'm'}^{00}.
\end{equation}
ACW \cite {Ackerman} found that
\begin{eqnarray}
\xi_{lm;l'm'}^{++}&=&-\delta_{m',m+2} \left[\delta_{l',l}{\sqrt{
(l^2-(m+1)^2)(l+m+2)(l-m)}\over (2l+3)(2l-1)}-{1 \over
2}\delta_{l',l+2}{\sqrt{ {(l+m+1)(l+m+2)(l+m+3)(l+m+4)\over
(2l+1)(2l+3)^2(2l+5)}}}\right. \nonumber \\
&& \left. -{1 \over 2}\delta_{l',l-2}{\sqrt{ {(l-m)(l-m-1)(l-m-2)(l-m-3)\over (2l+1)(2l-1)^2(2l-3)}}}\right] ,\nonumber \\
\xi_{lm;l'm'}^{--}&=&\xi_{l'm';lm}^{++},\nonumber \\
\xi_{lm;l'm'}^{+-}&=&{1 \over 2}\delta_{m',m}\left[-2 \, \delta_{l',l}
\frac{(-1+l+l^2+m^2)}{(2l-1)(2l+3)} \right. +
\delta_{l',l+2}\sqrt{\frac{((l+1)^2-m^2)((l+2)^2-m^2)}{(2l+1)(2l+3)^2(2l+5)}}
\nonumber \\
&& +\left.
\delta_{l',l-2}\sqrt{\frac{(l^2-m^2)((l-1)^2-m^2)}{(2l-3)(2l-1)^2(2l+1)}}\right],
\nonumber \\
\xi_{lm;l'm'}^{+0}&=&\frac{\delta_{m',m+1}}{\sqrt{2}}\left[\delta_{l',l}{(2m+1)\sqrt{(l+m+1)(l-m)} \over (2l-1)(2l+3)} \right. \nonumber \\
&&+ \left. \delta_{l',l+2}
\sqrt{\frac{((l+1)^2-m^2)(l+m+2)(l+m+3)}{(2l+1)(2l+3)^2(2l+5)}}
-\delta_{l',l-2}\sqrt{\frac{(l^2-m^2)(l-m-1)(l-m-2)}{(2l-3)(2l-1)^2(2l+1)}}
\right], \nonumber\\
\xi_{lm;l'm'}^{-0}&=&-\xi_{l'm';lm}^{+0},\nonumber \\
\xi_{lm;l'm'}^{00}&=&\delta_{m,m'} \left[ \delta_{l,l'} \frac{(2l^2+2l-2m^2-1)}{(2l-1)(2l+3)} +\delta_{l',l+2}
\sqrt{\frac{((l+1)^2-m^2)((l+2)^2-m^2)}{(2l+1)(2l+3)^2(2l+5)}}
\right. \nonumber \\
&&\left. +\delta_{l',l-2} \sqrt{\frac{(l^2-m^2)((l-1)^2-m^2)}{(2l-3)(2l-1)^2(2l+1))}} \right].
\label{thebigformula}
\end{eqnarray}
The term linear in $\zstar$ has already been evaluated before.
\be
\label{delta21}
\Delta_2^{(1)}(l,m;l',m')=i\int_0^{\infty} {\rm d}k ~k^2 P_{2}(k)\left(\Theta_{l'}(k){\partial \Theta_l(k) \over \partial k}-\Theta_l(k) {\partial \Theta_{l'}(k) \over \partial k}\right)\left(z_+\chi^{(a)+}_{lm;l'm'} +z_-\chi^{(a)-}_{lm;l'm'} +z_0\chi^{(a)0}_{lm;l'm'} \right)
\ee
where all $\chi^{(a)}$'s are given from Eq.~\eqref{chi} to \eqref{chi-}.

The term independent of $\zstar$ can be evaluated using the identity
\begin{eqnarray}
\sum_{i,j=1}^3\frac{k_ik_j}{k^2}Y_l^m(\hat{\bf k})\Theta_l(k)\frac{\partial }{\partial k_i}\frac{\partial }{\partial k_j}\pbra{Y_{l'}^{m'*}({\hat {\bf k}})\Theta_{l'}(k)}&=&\sum_{i,j=1}^3\frac{k_i}{k}Y_l^m(\hat{\bf k})\Theta_l(k)\frac{\partial }{\partial k_i}\pbra{\frac{k_j}{k}\frac{\partial}{\partial k_j}\pbra{Y_{l'}^{m'*}({\hat {\bf k}})\Theta_{l'}(k)}}\nn
&=&Y_{l}^{m}({\hat {\bf k}})\Theta_{l}(k)\hat{\bf k}\cdot\textrm{\boldmath $\nabla_k$}\bbra{\hat{\bf k}\cdot\textrm{\boldmath $\nabla_k$}\pbra{Y_{l'}^{m'*}({\hat {\bf k}})\Theta_{l'}(k)}}
\end{eqnarray}
From Eq.~\eqref{tricky1}, we know that
\be
\hat{\bf k}\cdot\textrm{\boldmath $\nabla_k$}\pbra{Y_{l'}^{m'*}({\hat {\bf k}})\Theta_{l'}(k)}=\frac{\partial \Theta_{l'}(k)}{\partial k}Y_{l'}^{m'*}({\hat {\bf k}})
\ee
and
\be
\hat{\bf k}\cdot\textrm{\boldmath $\nabla_k$}\bbra{\hat{\bf k}\cdot\textrm{\boldmath $\nabla_k$}\pbra{Y_{l'}^{m'*}({\hat {\bf k}})\Theta_{l'}(k)}}=\frac{\partial^2 \Theta_{l'}(k)}{\partial k^2}Y_{l'}^{m'*}({\hat {\bf k}})
\ee
These give
\be
\label{delta20}
\Delta_2^{(0)}(l,m;l',m')=\frac12 \delta_{l,l'}\delta_{m,m'}\int_0^{\infty} {\rm d}k ~k^2 P_{2}(k)\left[\pbra{\frac{\partial \Theta_l(k)}{\partial k}}^2-\Theta_l(k)\frac{\partial^2 \Theta_l(k)}{\partial k^2}\right]
\ee
To recap:  the modification of the correlations $\langle a_{lm}a_{l'm'}^* \rangle$ caused by the
violation of translational invariance is defined by Eq.~(\ref{Delta}).
It can be decomposed into two pieces, $\Delta_1(l,m,l',m')$ and $\Delta_2(l,m,l',m')$, and each can be expressed as three components depending on their
dependence on ${\bf z}_*$ in Eq.~\eqref{deltadecomp} and \eqref{deltadecomp2}.  The quadratic piece in $\Delta_1(l,m,l',m')$ is given
by (\ref{delta2}), the ${\bf z}_*$-independent piece by (\ref{delta0}), and the linear
piece by (\ref{linear}), whose terms are given by (\ref{termi}-\ref{linear end}). Meanwhile, the quadratic piece in $\Delta_2(l,m,l',m')$ is given by \eqref{delta22}, the linear piece by \eqref{delta21}, and the $\zstar$-independent piece by \eqref{delta20}.

While these expressions appear formidable, the good news is that coefficients at multipole
$l$ are only correlated with those at $l-2 \leq l' \leq l+2$.  The correlation matrix is sparse,
making the analysis of CMB data computationally tractable \cite{Groeneboom:2008fz}.

\section{Set up for A Special Line or Plane}

In this section we extend the results obtained for the case of a preferred point in space to the cases where translational invariance is broken by a special line or point. Since many of the steps are similar to the special point case we will be brief.

\begin{figure}[b]
\includegraphics[width=0.4\textwidth]{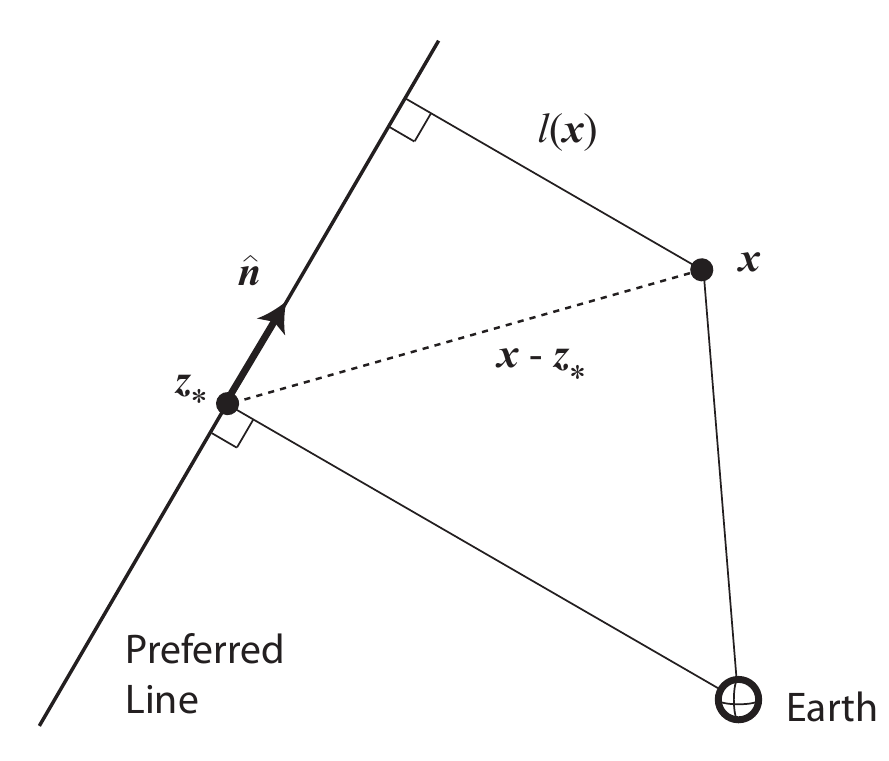} ~~~~~~~~
\includegraphics[width=0.4\textwidth]{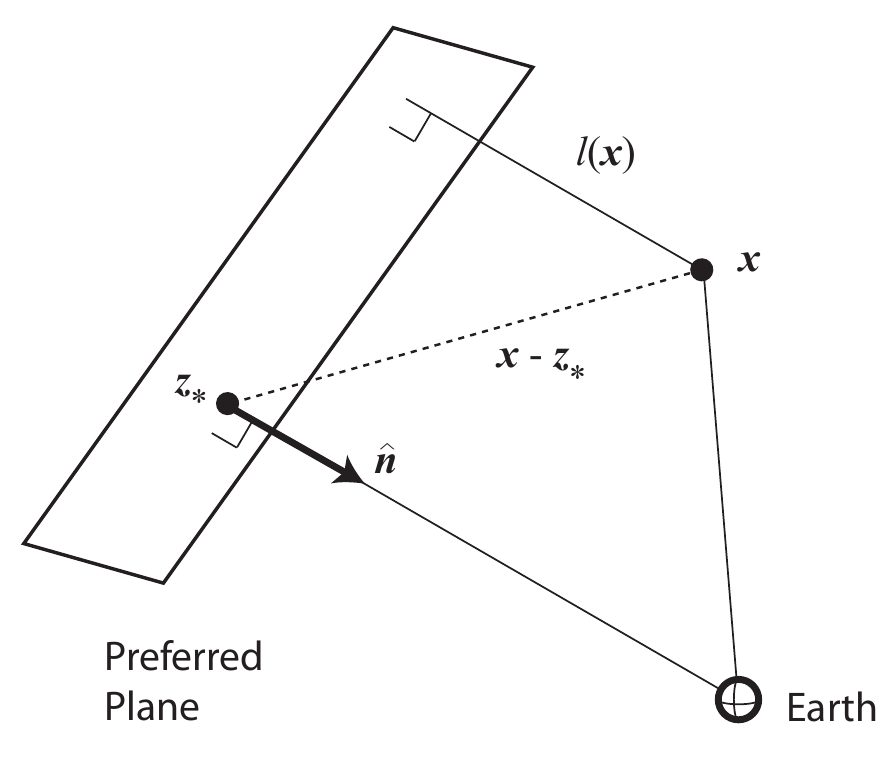}
\caption{A preferred line in space can be specified by its closest point, ${\bf z}_*$, and
a unit tangent vector $\hat {\bf n}$; a preferred plane can be specified by its closest point
and a unit normal vector.  The distance $l({\bf x})$ to any point ${\bf x}$ in space
is measured perpendicularly to the line or plane.}
\label{specialline}
\end{figure}

To specify the location of a preferred line in space requires a point ${\bf z}_*$ and a
unit tangent vector ${\bf n}$.  (Note that we place Earth at the center of our coordinate system,
so that the specification of any point defines a vector pointing from us to the point.)
Since any point on the line will do, without loss of generality we can take ${\bf z}_*$ to
be the point closest to us, implying the constraint ${\bf n} \cdot {\bf z}_*=0$. This is illustrated by the diagram on the left in Figure~\ref{specialline}.

In order to simplify the calculation, we first align the preferred direction with the $z$ axis. In that case, the rotational invariance about the
$z$ axis and the translational invariance along this preferred direction are left unbroken. These symmetries imply that the most general form of the two point correlation of energy density correlations is
\begin{equation}
 \langle \tilde \delta({\bf k})\tilde \delta({\bf q}) \rangle= \delta(k_z+q_z)e^{-i({\bf k}_\perp+{\bf q}_\perp)\cdot {\bf z}_*}P_t(k_\perp, q_\perp, k_z,{\bf k}_\perp \cdot {\bf q}_\perp),
\end{equation}
so that
\begin{eqnarray}
 \langle \delta ({\bf x})  \delta({\bf y}) \rangle &=&\int d^3 k \int  d^3 q~  e^{i{\bf k}\cdot {\bf x}}e^{i{\bf q}\cdot {\bf y}}\langle \tilde \delta({\bf k})\tilde \delta({\bf q}) \rangle \nn
&=&\int dk_z\int d^2 k_\perp \int d^2 q_\perp~ e^{ik_z(x_z-y_z)}e^{i{\bf k}_\perp\cdot ({\bf x}_\perp-{\bf z_{*\perp}})}e^{i{\bf q}_\perp\cdot ({\bf y}_\perp-{\bf z_{*\perp}})}P_t(k_\perp, q_\perp, k_z,{\bf k}_\perp \cdot {\bf q}_\perp)
\end{eqnarray}
with $P_t$ symmetric under the interchange of ${\bf k}_\perp$ and ${\bf q}_\perp$. Here we have decomposed the position and wave vectors along the $z$ axis and the two dimensional subspace perpendicular to that which is denoted by a subscript $\perp$. In the limit that there is no violations of translational (and rotational) invariance, $P_t(k_\perp, q_\perp, k_z,{\bf k}_\perp \cdot {\bf q})$ reduces to $P(k)\delta^2({\bf k}_\perp+{\bf q}_\perp)$, where $k=\sqrt{{\bf k_\perp}^2+k_z^2}$.  We now assume the violations of translational (and rotational) invariance are small and hence that $P_t$ is strongly peaked about ${\bf k_\perp}=-{\bf q_\perp}$.   We introduce the variables ${\bf p}_\perp={\bf k}_\perp+{\bf q}_\perp$, ${\bf l}_\perp=({\bf k}_\perp-{\bf q}_\perp)/2$ and follow the same steps in the point case. Then,
\begin{eqnarray}
\langle \delta ({\bf x})  \delta({\bf y}) \rangle =\int dk_z \int d^2 l_\perp~ e^{ik_z(x_z-y_z)}e^{i{\bf l}_\perp\cdot ({\bf x}_\perp-{\bf y}_\perp)}P_t(l_\perp,l_\perp, k_z, -l_\perp^2)\cdot\nn
\int d^2 p_\perp~e^{-A(l_\perp, k_z){p_\perp^2}/2-B(l_\perp, k_z)({\bf p}_\perp \cdot {\bf l}_\perp)^2 /(2l_\perp^2)} e^{i {\bf p}_\perp\cdot {\bf z}_\perp}\nn
\end{eqnarray}
where ${\bf z}_\perp=({\bf x}_\perp+{\bf y}_\perp-2{\bf z_{*\perp}})/2$. Performing the integral over $d^2 p_\perp$, we find that,
\begin{equation}
\langle \delta ({\bf x})  \delta({\bf y}) \rangle =\int dk_z\int d^2 l_\perp~ e^{ik_z(x_z-y_z)}e^{i{\bf l}_\perp\cdot ({\bf x}_\perp-{\bf y}_\perp)}P_t(l_\perp,l_\perp, k_z, -l_\perp^2)\sqrt{{{(2\pi)^2} \over {\rm det }C}}\left(1-{z_\perp^TC^{-1}z_\perp \over 2}+\ldots \right)
\end{equation}
where $C_{ij}=A(l_\perp, k_z)\delta_{ij}+B(l_\perp, k_z)\dfrac{l_{\perp i} l_{\perp j}}{l_\perp^2}$ is a $2\times2$ matrix, ${\rm det}C=A^2+AB$, and
\be
C^{-1}_{ij}={1 \over A}\delta_{ij}-{B \over A(A+B)}{l_{\perp i}l_{\perp j} \over l_\perp^2}
\ee
We can define
\be
P(l_\perp, k_z)=\sqrt{{{(2\pi)^2} \over {\rm det }C}}P_t(l_\perp,l_\perp, k_z -l_\perp^2)
\ee
and plug in the expression of $C_{ij}^{-1}$ in terms of $A(l_\perp,k_z)$ and $B(l_\perp, k_z)$. This gives after relabeling, ${\bf l}_\perp  \rightarrow  {\bf k}_\perp$
\begin{equation}
\label{generalline}
\langle \delta ({\bf x})  \delta({\bf y}) \rangle =\int d^3k~ e^{i{\bf k} \cdot ({\bf x}-{\bf y})} P(k_\perp, k_z) \bbra{1-\frac{z_\perp^2}{2A}+\frac{B}{2A(A+B)}\frac{({\bf k}_\perp\cdot{\bf z}_\perp)^2}{k_\perp^2}}
\end{equation}
Note that we want the leading term in the expansion in $z$ to correspond to the standard cosmology and hence $P(k_\perp, k_z) =P(k)$, where $k=\sqrt{k_\perp^2 +k_z^2}$.
Finally, to make the preferred direction arbitrary,  we  replace all position vectors $a_z$ with ${\bf n}\cdot{\bf a}$ and also replacing ${\bf a}_\perp$ with ${\bf a}-{\bf n}({\bf n}\cdot{\bf a})$ in Eq.~\eqref{generalline}.

As in the special point case we note that another way to get a small violation of translational is if there is a small parameter $\epsilon$ and $ P_t$ takes the form,
\be
 P_t(k_\perp, q_\perp, k_z,{\bf k}_\perp \cdot {\bf q}_\perp)={c \over k^3}\delta({\bf k}+{\bf q}) +\epsilon P'_t(k_\perp, q_\perp, k_z,{\bf k}_\perp \cdot {\bf q}_\perp)
\ee
where $P'_t$ cannot be expanded in any simple way. This is what happened in Ref.~(\cite{Tseng:2009xw}).

A preferred plane can be specified by a point ${\bf z}_*$ and a
unit normal vector ${\bf n}$. We can again choose ${\bf z}_*$ to be the point on the plane closest to us, implying
a constraint ${\bf n \times} {\bf z}_*=0$, as shown on the right-hand side of Figure~\ref{specialline}. Notice that the rotational invariance about the ${\bf n}$ axis and the translational invariance along the ${\bf n}$ direction are unbroken. These symmetries imply
\begin{equation}
 \langle \tilde \delta({\bf k})\tilde \delta({\bf q}) \rangle= \delta^2({\bf k}_{\|}+{\bf q}_{\|})e^{-i(k_n+q_n)z_{*n}}P_t(k_{\|}, k_n, q_n)
\end{equation} 
so that
\begin{eqnarray}
 \langle \delta ({\bf x})  \delta({\bf y}) \rangle &=&\int d^3 k \int  d^3 q~  e^{i{\bf k}\cdot {\bf x}}e^{i{\bf q}\cdot {\bf y}}\langle \tilde \delta({\bf k})\tilde \delta({\bf q}) \rangle \nn
&=&\int d^2k_{\|} \int d k_n  \int d q_n~ e^{i{\bf k}_{\|}\cdot({\bf x}_{\|}-{\bf y}_{\|})}e^{ik_n(x_n-z_{*n})}e^{iq_n(y_n-z_{*n})}P_t(k_{\|}, k_n, q_n)
\end{eqnarray}
Here we have decomposed the position and wave vectors along the normal vector ${\bf n}$ and the two dimensional subspace parallel to the plane which is denoted by a subscript $\|$. Then we change variables $p_n=k_n+q_n$, $l_n=(k_n-q_n)/2$ and perform the integral over $dp_n$ to get
\begin{equation}
\langle \delta ({\bf x})  \delta({\bf y}) \rangle =\int d^2k_{\|} \int d l_n ~ e^{i{\bf k}_{\|}\cdot({\bf x}_{\|}-{\bf y}_{\|})}e^{il_n(x_n-y_n)}P_t(k_{\|},l_n,l_n)\sqrt{{{2\pi} \over A}} \left(1-\frac{z_n^2}{2A}+\ldots \right)
\end{equation}
After relabeling $l_n\rightarrow k_n$ and defining
\be
P(k_{\|},l_n)=\sqrt{{{2\pi} \over A}}P_t(k_{\|},l_n,l_n)
\ee
we have
\begin{equation}
\label{generalplane}
\langle \delta ({\bf x})  \delta({\bf y}) \rangle =\int d^3k~ e^{i{\bf k} \cdot ({\bf x}-{\bf y})} P(k_{\|}, k_n) \bbra{1-\frac{z_n^2}{2A}}
\end{equation}
Finally, for the reason that we want the leading order term to correspond to the standard cosmology, we replace $P(k_{\|}, k_n)$ with $P(k)$, where $k=\sqrt{k_{\|}^2+k_n^2}$. 

\section{Conclusions}

We have investigated the observational consequences of a small violation of translational
invariance on the temperature anisotropies in the cosmic microwave background.  Three cases were
investigated, based on the assumption of a preferred point, line, or plane in space,
and a quadratic dependence on the distance to the preferred locus of points.
Explicit formulae were presented for the correlations $\langle a_{lm} a^*_{l'm'}\rangle$
between spherical harmonic coefficients of the CMB temperature field in the case of a special point. The expressions we have derived may be used to directly compare CMB observations
against the hypothesis of perfect translational invariance during the inflationary era, as
part of a systematic framework for constraining deviations from the standard
paradigm of primordial perturbations.
Explicit expressions for the correlations  $\langle a_{lm} a^*_{l'm'}\rangle$ can also be derived for the special line and plane cases.

One can also test the hypothesis of perfect translational invariance during the inflationary era using data on the large scale distribution of galaxies and clusters of galaxies,
using, in the special point case,
\be
\label{ptgen1a}
\langle \delta ({\bf x})  \delta({\bf y}) \rangle =\int {d^3 l \over (2 \pi)^3} e^{i{\bf l}\cdot ({\bf x}-{\bf y})}P_0(l)+\frac{({\bf x}+{\bf y}-2{\bf z_*})^2}{4}\int {d^3 l \over (2 \pi)^3} e^{i{\bf l}\cdot ({\bf x}-{\bf y})}P_1(l)+\int {d^3 l \over (2 \pi)^3} e^{i{\bf l}\cdot ({\bf x}-{\bf y})}P_2(l)\frac{\bbra{\bf l\cdot({\bf x}+{\bf y}-2{\bf z_*})}^2}{4l^2}.
\ee
The work in Section II  suggests that $P_1(k)$ and $P_2(k)$ are proportional to $P_0(k)$ and so the corrections to the microwave background anisotropy and the large scale distribution of galaxies are characterized by five parameters, two are these constants of proportionality and three are the parameters to specify the sepcial point including the direction and the magnitude of $\zstar$.

\section*{Acknowledgments}

This research was supported in part by the U.S. Department of Energy and by the Gordon and Betty Moore Foundation.

\end{document}